\documentclass{ws-ijmpa}
\usepackage{times,xcolor,amsmath,amssymb,epsfig,graphics,graphicx,multirow}
\usepackage[super,compress]{cite}
\newcommand{\be}{\begin{eqnarray}}
\newcommand{\ee}{\end{eqnarray}}

\newcommand{\er}{$\pm$}
\newcommand{\beq}{\begin{eqnarray}}
\newcommand{\eeq}{\end{eqnarray}}
\newcommand{\bc}{\begin{center}}
\newcommand{\ec}{\end{center}}
\newcommand{\bit}{\begin{itemize}}
\newcommand{\eit}{\end{itemize}}

\definecolor{lightyellow}{cmyk}{0,0,0.5,0}
\definecolor{lightred}{rgb}{1,0.5,0.5}
\definecolor{lightgreen}{rgb}{0,0.4,0}
\definecolor{lightblue}{rgb}{0.5,0.5,1}
\definecolor{darkred}{rgb}{0.8,0,0}
\definecolor{darkgreen}{rgb}{0,0.4,0}
\definecolor{darkcyan}{cmyk}{1,0.3,0.3,0.3}
\definecolor{darkblue}{rgb}{0,0,0.6}
\definecolor{lightbrown}{rgb}{0.7,0.3,0.3}
\definecolor{darkbrown}{rgb}{0.5,0,0}
\begin{document}
\markboth{A.V.~Sarantsev and E. Klempt}{Scalar and tensor mesons in $d\bar d$, $s\bar s$ and  $gg\to f_0, f_2$}

%
\catchline{}{}{}{}{}
%

\title{\bf\boldmath Scalar and tensor mesons in $d\bar d$, $s\bar s$ and  $gg\to f_0, f_2$}

\author{A.V.~Sarantsev and E. Klempt}

\address{Helmholtz--Institut f\"ur Strahlen- und Kernphysik, Universit\"at Bonn, 53115 Bonn, Germany}
\maketitle

\begin{history}
\received{\today}
\end{history}

\begin{abstract}
LHCb data on $B^0\to J/\psi  \pi^+\pi^-$,  $B_{s} ^0\to J/\psi  \pi^+\pi^-$ and  $B_{s} ^0\to J/\psi  K^+K^-$
are fitted in a coupled-channel analysis jointly with data on radiative $ J/\psi $ decays,
$\pi\pi$ elastic scattering and further data sets. Branching ratios for $B_{(s)} ^0\to J/\psi f_0,f_2$
are determined and  compared to those from radiative $J/\psi$ decays.
Above 1500\,MeV, only little intensity is observed in
$B_{(s)} ^0\to J/\psi  f_0,f_2$ decays. In radiative $ J/\psi $ decays, the intensities are strong and
peak at 1865\,MeV in the scalar wave and at 2210\,MeV in the tensor wave. This
pattern is interpreted as further support for the glueball
interpretation of $G_0(1865)$ and of $G_2(2210)$.
\end{abstract}

\ccode{PACS numbers: 25.75.-q}

\section{Introduction}
Scalar mesons, the excited states of the QCD vacuum, are intriguing objects. In the low-energy
region, a nonet of controversially discussed scalar mesons is observed. These mesons could
be four-quark states~\cite{Jaffe:1976ig}, molecules~\cite{vanBeveren:1986ea}
or could have a $q\bar q$ seed that acquires a meson cloud but still forms the ground states
of the meson spectrum~\cite{Klempt:2021nuf}. Above 1\,GeV,
a series of scalar isoscalar resonances is seen~\cite{ParticleDataGroup:2022pth} which are  believed to house
the scalar glueball, a state
without constituent quarks. Probably, the glueball mixes with scalar mesons with similar mass.
Often, the three mesons $f_0(1370)$, $f_0(1500)$ and $f_0(1710)$ are thought to result
from the mixing of the expected $(u\bar u+d\bar d)/\sqrt2$ and $s\bar s$ states with the scalar glueball
\cite{Amsler:1995tu,Amsler:1995td}. A large number of meson-glueball mixing scenarios
have been proposed using these three mesons, all assuming that the sum of the glueball fractions
in these three mesons should add up to one. A list of references can be found
elsewhere~\cite{Klempt:2021wpg}.

\begin{figure}[t!]
\centering
\includegraphics[width=0.98\textwidth]{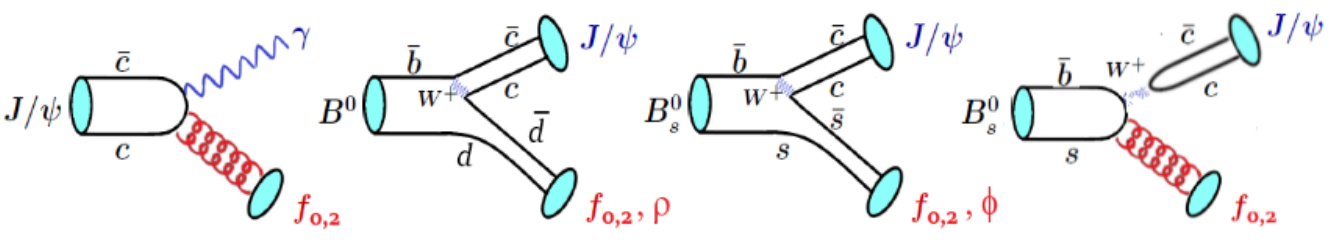}\\[-4ex]
\hspace{-10mm}a\hspace{30mm}b\hspace{30mm}c\hspace{30mm}d
\caption{In radiative $J/\psi$ decays, the $c\bar c$ pair emits a photon and
converts into two gluons. These may interact forming a glueball (a). In $B^0$ (or $\bar B^0$) decays
into a $J/\psi$ and a neutral meson, the $\bar b$ (or $b$) quark can decay into a $c$ (or $\bar c$)
quark and creates a $\bar c d$ (or $c\bar d)$ pair. With the $\bar d$ (or $d)$ quark,
a light-quark neutral meson is formed (b). In the case of $B_s^0$ ($\bar B_s^0$) decays,
a $s\bar s$ meson recoils against the $J/\psi$ (c).
Alternatively, a $W^+$ can be exchanged and two gluons can be radiated (d). 
If this process would dominate the reaction, 
the ratio of frequencies for the production of scalar mesons would be similar to those
for radiative $J/\psi$ decays.}
\label{Reactions}
\end{figure}

The ideal path to search for glueballs is the study of radiative decays of $J/\psi$ mesons.
 In these decays, the $J/\psi$ resonance emits a photon and the $c\bar c$ 
system converts into two gluons which should interact to form glueballs
(see Fig.~\ref{Reactions}, left). 
Recently, high-precision data on radiative $J/\psi$ decays into $\pi^0\pi^0$ and
$K_SK_S$, decomposed into partial waves, have became available~\cite{BESIII:2015rug,BESIII:2018ubj}.

Fits to these data were presented in Refs.~\cite{Sarantsev:2021ein,Rodas:2021tyb}.
In the latter publication, the data above 1\,GeV were fit by four scalar and 
three tensor mesons. The $S$-wave
required a ``back\-ground" amplitude due to $\pi^0\gamma$ ($K_S\gamma$) resonances 
recoiling against a $\pi^0$ ($K_S$) with poles outside of the kinematic accessible 
region; see Section 8.4.9 \cite{Klempt:2022ipu} in  
Ref.~\cite{Gross:2022hyw} for a discussion.
In Ref.~\cite{Sarantsev:2021ein}, the full energy
range was fitted, and the fit was constrained by a large set of additional data,
the $\eta\eta$ and $\omega\phi$ $S$-wave from radiative $J/\psi$ decays
 \cite{Ablikim:2013hq,Ablikim:2012ft},  the CERN-Munich~\cite{Grayer:1974cr} data and the
$K_{\rm e4}$~\cite{Batley:2010zza} data. Further, data were used from
BNL and GAMS and 15 Dalitz plots for different reactions
from $\bar pN$ annihilation at rest (see Ref.~\cite{Sarantsev:2021ein} for references).
In that analysis, ten scalar mesons were identified. The masses and widths of four resonances 
were compatible with those found in Ref.~\cite{Rodas:2021tyb} but each of the four resonances is accompanied by a resonances at slightly lower mass.
Due to the wider range at low energies, $f_0(500)$ and $f_0(980)$ were observed in~\cite{Sarantsev:2021ein} additionally. 
The yield of scalar mesons showed a strong peak, with a mass and width determined to be
$M=1865$\er25$^{+10}_{-30}$ {\rm MeV} and $\Gamma= 370$\er$50^{+30}_{-20}$ {\rm MeV}, its yield in radiative $J/\psi$ decays to ($5.8\pm 1.0)\,10^{-3}$. The peak was interpreted as
evidence for the scalar glueball. 

SU(3) relates the decay modes of meson resonances to their internal structure;
see Fig. 63.2 in~\cite{ParticleDataGroup:2022pth}.
An analysis of the decay modes of the scalar mesons into $\pi\pi$, $K\bar K$, $\eta\eta$ and $\eta\eta'$ 
shows~\cite{Klempt:2021wpg} that the mesons above 1\,GeV can be classified as
mainly singlet or mainly octet mesons in SU(3). The mesons reported in Ref.~\cite{Rodas:2021tyb}
were shown to be mainly octet mesons, the mesons seen additionally in Ref.~\cite{Sarantsev:2021ein}
have a mainly-singlet structure. 

At first, this is surprising: Octet mesons
should not be produced in radiative $J/\psi$ decays. However, allowing for a small glueball fraction
in the wave function of the scalar mesons improved the fits to the decay branching ratios for several scalar mesons. 
This suggests that the scalar glueball is spread across many scalar mesons, and the production of high-mass scalar mesons in radiative $J/\psi$ decays is due to their glueball component in the wave function. In fact, the glueball content is maximized at 1865 MeV, and the sum of the observed glueball contributions in six mesons adds up to (78\er 18)\%~\cite{Klempt:2021wpg}.
Note that in this analysis, unlike all previous mixing scenarios, the existence of a scalar glueball is not imposed but deduced from the decay modes.

Scalar mesons with a large singlet component, like $f_0(500)$ and $f_0(1370)$, can be (and are) produced in radiative $J/\psi$ decays even if they have no glueball component. Above 1500 MeV, singlet and octet scalar mesons are produced at similar rates. Thus, the $q\bar q$
components of the produced mesons do not seem to play a significant role in radiative $J/\psi$ decays. In high-mass (radially excited) scalar mesons, the
$q\bar q$ component has little overlap with the initial $gg$
pair. The production could be driven by the gluon-gluon component in the mesonic wave function.
 
The tensor mesons $f_2(1270)$ and $f_2'(1525)$ are produced as well but above these mesons, 
only a small enhancement at 2210\,MeV was observed~\cite{Klempt:2022qjf}. This enhancement was discussed as a candidate
for the lowest-mass tensor glueball mixed with ordinary tensor mesons expected in this mass region.
 In Ref.~\cite{Vereijken:2023jor},
$f_2(1950)$ is argued to be a better glueball candidate, although the mixing of ordinary tensor mesons and the tensor glueball was not discussed.

In this paper, we test the hypothesis that high-mass mesons in radiative $J/\psi$ decays are 
produced predominantly due to their glueball content. We do this by comparing the yields of scalar and tensor mesons in 
radiative $J/\psi$ decays with those in the decays $ B^0\to J/\psi f_{0,2}$ and $ B^0_{s}\to J/\psi f_{0,2}$.
Figure~\ref{Reactions}, center and right, visualizes these decays.   Both decay processes proceed under similar kinematical 
conditions. In the initial state, a $d\bar d$ or $s\bar s$ pair is formed. 
The produced $q\bar q$ pair may have a large invariant mass, e.g. 1900\,MeV, but then
its coupling to $q\bar q$ mesons could be suppressed by the small
density of the mesonic wave function at the origin. 
The reaction $ B^0_{s}\to  \eta_c f_{0,2}$ would also be
well suited if the angular distributions of $\pi\pi$ and $K\bar K$ were available. 

\section{LHCb data on $B^0_{s}\to J/\psi \pi^+\pi^-$ and $J/\psi K^+K^-$}

$B^0$ and $B^0_{s}$ mesons, collectively denoted as $B^0_{(s)}$, mix with their antiparticles $\bar{B}^0_{(s)}$. This mixing provides access to decay-time-dependent $CP$ asymmetries and offers opportunities to discover physics beyond the Standard Model \cite{Bigi:2000yz}. The LHCb collaboration has extensively studied reactions such as $\bar{B}^0_{(s)} \leftrightarrow B^0_{(s)} \to J/\psi$ plus a light-quark meson, where the meson subsequently decays into $\pi\pi$ or $K\bar{K}$ \cite{LHCb:2012ad,LHCb:2012ae,LHCb:2013dkk,LHCb:2013kpp,LHCb:2013odx,LHCb:2013eus,LHCb:2014iah,LHCb:2014vbo,LHCb:2014ooi,LHCb:2014tol,LHCb:2014xpr,LHCb:2017hbp,LHCb:2019sgv,LHCb:2019nin}. In several publications, the LHCb collaboration presented spherical harmonic moments and their dependence on the $\pi^+\pi^-$ or $K^+K^-$ invariant mass, interpreting these results in terms of contributing resonances.

Here, we incorporate LHCb data on $\bar{B}^0_{(s)} \leftrightarrow B^0_{(s)} \to J/\psi$ plus $\pi^+\pi^-$ or $K^+K^-$ into a coupled-channel analysis with data on radiative $J/\psi$ decays into $\pi^0\pi^0$, $K_s^0 K_s^0$, $\eta\eta$, $\phi\omega$ and other data discussed previously \cite{Sarantsev:2021ein}. Based on the evidence and properties of states reported in Ref.~\cite{Sarantsev:2021ein}, we can estimate the branching ratios of resonances weakly produced in $\bar{B}^0_{(s)} \leftrightarrow B^0_{(s)} \to J/\psi f_{0,2}$ decays.

The decays $B^0_{(s)}\to J/\psi\,\pi^+\pi^-$ or $J/\psi\,K^+K^-$ proceed via seven helicity amplitudes
(one scalar, three vector, three tensor waves).
These amplitudes were determined by analyzing the distribution of the angle between the decay planes
of  $J/\psi\to\mu^+\mu^-$ and  $\pi^+\pi^-$ in the rest frame of $\bar B^0_{(s)}$ or $B^0_{(s)}$. 
The $\pi^+\pi^-$ or $K^+K^-$ pairs were described by resonant or
nonresonant amplitudes. The data were visualized by presenting the
mass dependence of the spherical harmonic moments of the $\cos\theta_{h\bar h}$
distributions, where $h$ stand for $\pi^+$ or $K^+$.

The spherical harmonic moments $\langle Y^0_l\rangle$ were presented as efficiency-corrected and background-subtracted 
invariant mass distributions weighted by orthogonal and normalized spherical harmonic functions
given as $Y^0_l(\cos\theta_{h\bar h})$.
In most cases, spherical harmonic moments $\langle Y^0_0\rangle$, $\langle Y^1_l\rangle$, $\cdots$
$\langle Y^0_6\rangle$ were provided. They contain information on the $S$, $P$, and $D$ waves, 
as well as interference between different partial waves. The odd moments are nearly negligible.
Specifically, 
$\langle Y^0_0\rangle$ represents the event distribution, $\langle Y^0_2\rangle$ represents
the sum of the $P$ and $D$ waves and the $S-D$ interference, and $\langle Y^0_4\rangle$ represents 
the $D$ wave.

\begin{figure}[pt]
\centering
\vspace{-1mm}
\begin{tabular}{cc}
\raisebox{3mm}{\includegraphics[height=0.25\textwidth,width=0.35\textwidth]{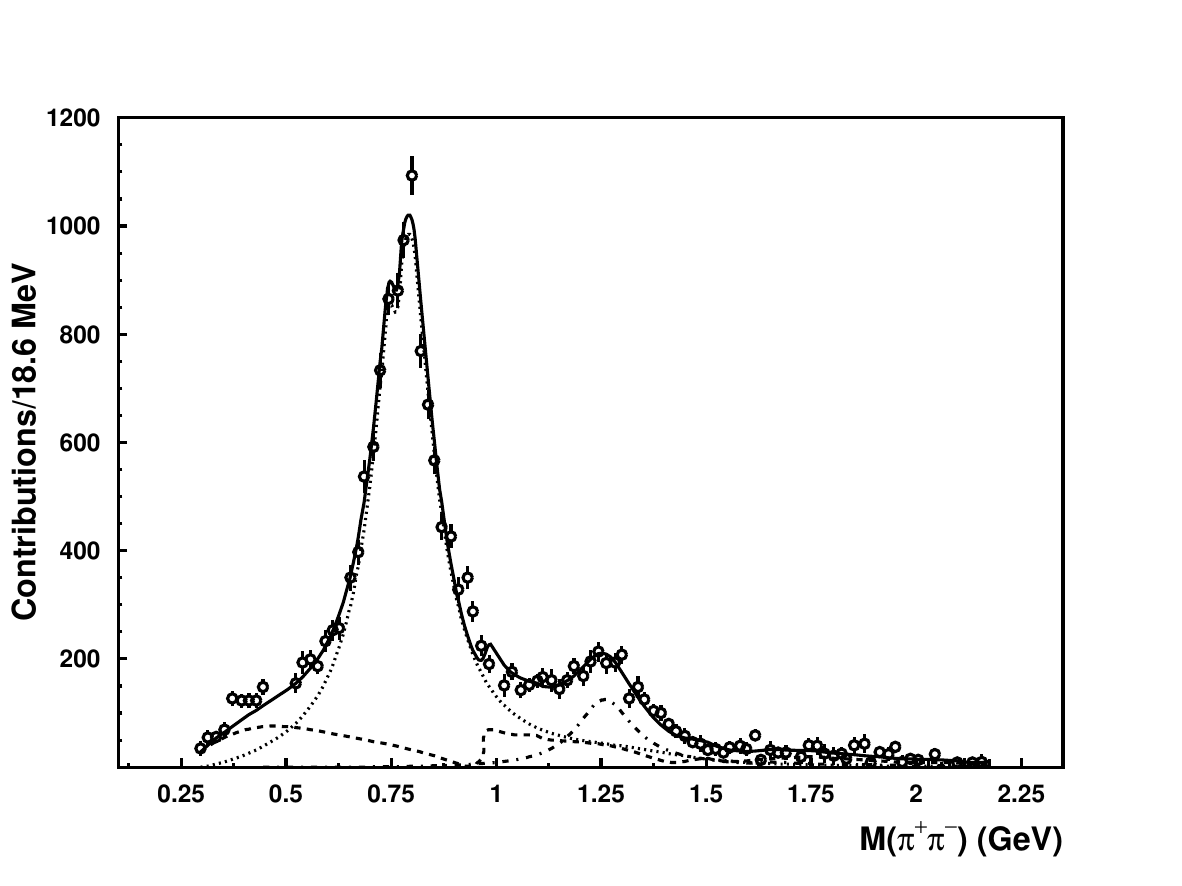}}&
\hspace{-6mm}\includegraphics[height=0.25\textwidth,width=0.65\textwidth]{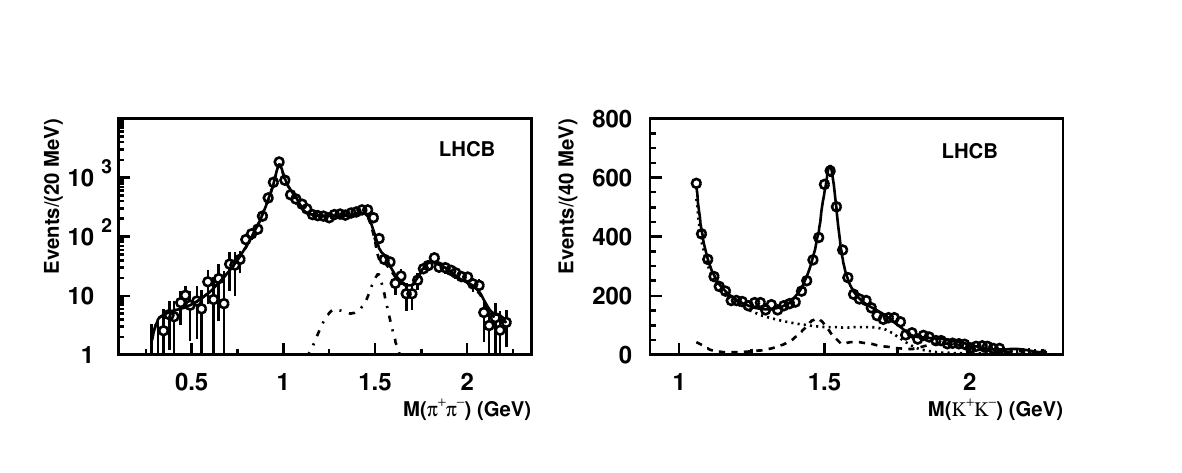}
\vspace{-3mm}
\end{tabular}
\caption{\label{bsdecays-tot}Left: The $\pi^+\pi^-$ invariant mass from the reaction
$\bar B^0\leftrightarrow B^0\to J/\psi \pi^+\pi^-$ \cite{LHCb:2014vbo}.
The solid curve represents our fit described below, including partial wave
contributions: $J^{PC}=2^{++}$ (dashed dotted), $1^{--}$ (dotted),
$0^{++}$ (dashed). Center and right: the $\pi^+\pi^-$ and $K^+K^-$
invariant mass distribution from reaction $\bar
B^0_s\leftrightarrow B^0_s\to J/\psi \pi^+\pi^-$ \cite{LHCb:2014ooi}
and $ J/\psi K^+K^-$ \cite{LHCb:2017hbp}.
 The solid curve represents the total contribution of our fit described in the text.  Partial wave
contributions are represented by the dashed dotted ($J^{PC}=2^{++}$), 
 dotted curve ($1^{--}$), and dashed curve ($0^{++}$). \vspace{-15mm}}
\begin{center}
\includegraphics[width=1.1\textwidth]{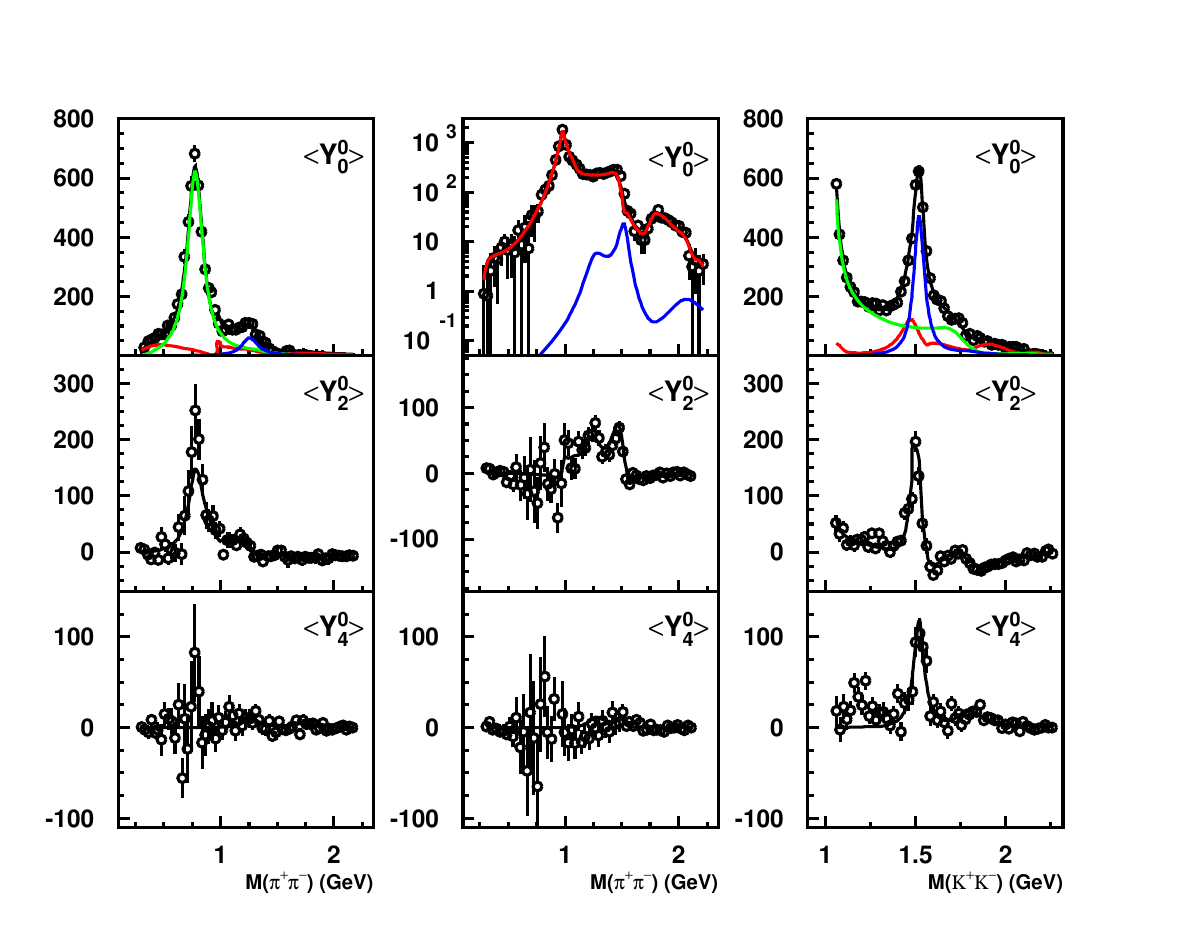}
\vspace{-11mm}
\end{center}
\caption{\label{bsdecays}The mass dependence of the spherical harmonic moments of
$\cos \theta_{h\bar h}$ for reactions $\bar B^0\leftrightarrow B^0\to J/\psi \pi^+\pi^-$ (left,  \cite{LHCb:2014vbo}),
$\bar B^0_{s} \leftrightarrow B^0_{s}\to J/\psi \pi^+\pi^-$
(center, \cite{LHCb:2014ooi}) and into $J/\psi\,K^+K^-$ (right, \cite{LHCb:2017hbp}). The points with error
bars represent the LHCb data points, while the solid black curves depict
our fit to $\langle Y^0_0\rangle$, $\langle Y^0_2\rangle$, $\langle Y^0_4\rangle$,
and to the background subtracted absolute yields. 
Odd moments and those with $L\geq 6 $ are
not included in the fit as they are not required by the data. The $S$-wave  ($P$-wave, $D$-wave)
contributions are shown as red (green, blue) curves. 
}
\end{figure}

Figure~\ref{bsdecays} (left) displays the $\pi^+\pi^-$ invariant mass distribution
from reaction
$\bar B^0\leftrightarrow B^0\to J/\psi \pi^+\pi^-$ \cite{LHCb:2014vbo}. The spectrum is predominantly influenced by the production of $\rho$ mesons, with a minor asymmetry attributed to $\rho$-$\omega$ interference. In the tensor wave, the presence of $f_2(1270)$ is observed. The scalar wave includes the $f_0(500)$. The $f_0(980)$ is identified from the fit to the moments, and the scalar intensity rises sharply at approximately 1 GeV. Above this energy, the intensity diminishes without significant structure.

In Ref.~\cite{LHCb:2014ooi}, $\bar B^0_{s} \leftrightarrow B^0_{s}$ mixing
and decays into the $J/\psi\pi^+\pi^-$ final state were investigated. The  $\pi^+\pi^-$
invariant mass distribution is depicted in Fig.~\ref{bsdecays-tot}, left. Five resonances were
identified: $f_0(980)$, $f_0(1500)$, $f_0(1770)$,
$f_2(1270)$, and $f_2'(1525)$. The production of tensor mesons is relatively  weak: the
tensor mesons are nearly ideally mixed, with
$f_2(1270)$ coupling minimally to the initial
$s\bar s$ pair and $f_2'(1525)$ having a small $\pi\pi$ branching ratio (BR). The $\pi^+\pi^-$ invariant mass spectrum is dominated by the $f_0(980)$ resonance, which exhibits 
strong coupling to both $s\bar s$ and $d\bar d$ quark pairs. 
In addition, clear observations were made for $f_0(1500)$ and $f_0(1770)$. 
It is evident that both resonances are not ideally mixed and couple to both $n\bar{n}$ and $s\bar{s}$ quark pairs.

The $J/\psi K^+K^-$ final state produced in $\bar B^0_{s}$ and $B^0_{s}$
decays was studied in Refs.~\cite{LHCb:2013kpp,LHCb:2017hbp}.
The invariant mass spectrum (see Fig.~\ref{bsdecays-tot}, right)
is dominated by the $\phi(1020)$ meson, with a small contribution observed
from $\phi(1680)$. The $S$-wave contributes $\sim$10\% to the data, although specific 
contributions from individual resonances are not detailed.

Four tensor mesons were identified: a notably significant $f_2'(1525)$, alongside contributions from
$f_2(1270)$, $f_2(1750)$ and
$f_2(1950)$. The data of Ref.~\cite{LHCb:2013kpp} were based on an
integrated luminosity of 1.0\,fb$^{-1}$,  Ref.~\cite{LHCb:2017hbp} utilized 3.0 fb$^{-1}$, encompassing the data of the previous publication. Therefore, the results from Ref.~\cite{LHCb:2013kpp} 
are not separately considered here.

In $B^0$ decays into $J/\psi$ and light-quark mesons, the $b$-quark transitions into a $\bar{c}$-quark through the emission of a $W^+$ boson, which subsequently decays into a $c$-quark and a $\bar{d}$-quark, forming the $J/\psi$ and a primary $d\bar{d}$ pair. Compared to $W^+ \to c\bar{s}$ transitions, $W^+ \to c\bar{d}$ transitions are highly suppressed due to the small CKM matrix element.

The $\bar{B}^0 \leftrightarrow B^0$ mixing decays into $J/\psi,\pi^+\pi^-$ were analyzed in~\cite{LHCb:2014vbo}. Six interfering resonances were identified: The $\pi^+\pi^-$ invariant mass distribution is predominantly influenced by the $\rho(770)$ resonance with some interference from $\omega$. The $f_2(1270)$ resonance is clearly discernible, and the fitting procedure also identifies $f_0(500)$, $\rho(1450)$, and $\rho(1700)$ as additional resonances.
Figure~\ref{bsdecays} displays the spherical harmonic moments for the three reactions.

In Ref.~\cite{Ropertz:2018stk}, the scalar pion form factor was derived from the decay channel $B^0_{s}\to J/\psi \pi^+\pi^-$. The resonances $f_0(500)$, $f_0(980)$, and $f_0(1500)$ were clearly observed contributing to the form factor, while $f_0(1710)/f_0(1770)$ were not observed, but a resonance at $f_0(2020)$ was identified at higher masses. The primary focus of the paper, however, was not to quantitatively compare branching ratios from radiative $J/\psi$ decays with the scalar form factor.
Here, our objective is to determine the branching ratios of $B^0\to J/\psi f_{0,2}$ and $B^0_{s}\to J/\psi f_{0,2}$, where $f_{0,2}$ subsequently decay into $\pi\pi$ or $K\bar{K}$, and to compare these ratios with the yields observed in radiative $J/\psi$ decays.

\section{BnGa fits to $B^0_{s}\to J/\psi \pi^+\pi^-$ and $J/\psi K^+K^-$}

In our analyses, the properties of scalar and tensor resonances, along with their respective numbers, are exclusively determined by non-LHCb datasets, which are not presented here. This approach allows us to test many more resonances for their potential contributions to the LHCb data.

For vector mesons, we adopt the resonances identified by the LHCb collaboration: $\rho(770)$, $\omega(782)$, and $\rho(1700)$ for $\bar{B}^0_{s} \leftrightarrow B^0_{s}\to J/\psi \pi^+\pi^-$, and $\phi(1020)$ and $\phi(1680)$ for $\bar{B}^0_{s} \leftrightarrow B^0_{s}\to J/\psi K^+K^-$. The fit provides fractional contributions of these resonances to the LHCb data, while absolute yields are determined by normalizing to the branching ratios (BRs) of $B^0\to J/\psi +\rho$ with a BR of $(2.55^{+0.18}_{-0.16})\cdot 10^{-5}$, and $B^0{s}\to J/\psi +\pi\pi$ with a BR of $(26\pm 6)\cdot 10^{-5}$.

Table~\ref{decays} presents our results compared to those of the LHCb collaboration. For mesons with lower masses, our findings are in good agreement with LHCb's results within similar uncertainties. Mesons with masses above 1.8,GeV, not reported by LHCb, are found to have small contributions in our analysis.

A slight discrepancy is observed around 1300\,MeV between the data and our fit in the $Y^0_4$ moment for $B^0_{s}\to J/\psi +\pi\pi$. In the LHCb analysis \cite{LHCb:2017hbp}, this enhancement is attributed to a $f_2(1270)\to K\bar{K}$ contribution. However, our coupled-channel analysis suggests that accommodating this enhancement with a larger $f_2(1270)\to K\bar{K}$ branching ratio is not warranted by the data; thus, we do not interpret this structure as evidence for additional physics.

\begin{table}[h!]
\tbl{\label{decays}Fractional contribution of resonances to $B^0$ and $B^0_{s}$ decays
into $J/\psi +\pi^+\pi^-$ and $J/\psi +K^+K^-$ in units of
$10^{-5}$. The BRs are normalized to the BR for
$B^0\to J/\psi +\rho$ (shown with a *) and for $B^0_{s}\to J/\psi +\pi\pi$ of
(26\er 6)$\cdot 10^{-5}$. The results of the LHCb collaboration are given as small numbers.
When no small numbers are given, the resonance was not considered in the LHCb fits.
} 
{
\centering
\renewcommand{\arraystretch}{1.4}
\begin{tabular}{cccccc}
\hline\hline
&&\multicolumn{1}{c}{$B^0$ decays}&& \multicolumn{2}{c}{$B^0_{s}$ decays}\\
&& $\pi^+\pi^-$&&$\pi^+\pi^-$&$K^+K^-$\\\hline
$f_0(500)$      &&0.3$^{+0.3}_{-0.1}$&&       $<2$   &   0    \\[-1.ex]
                     &&\tiny0.88$^{+1.2}_{-1.6}$ &        &          \\ \hline
$f_0(980)$      &&0.05\er0.03&&   15.0\er1.5      &  3.7\er0.4      \\[-1.5ex]
                     &&\tiny$<0.11$&& \tiny 12.8\er1.8        &          \\[-0.5ex] \hline
$f_0(1370)$    &&$<0.01$&& 5.0\er1.2      &0.8\er0.2         \\[-1.2ex]
                     &&&&\tiny $4.5^{+0.7}_{-4.0}$        & \\[-0.5ex] \hline
$f_0(1500)$    &&$<0.01$&& 3.0\er 1.0       &0.7\er0.3        \\ [-1.2ex]
                     &&&&\tiny $2.11^{+0.40}_{-0.29}$       &  \\[-0.5ex] \hline
$f_0(1710)$    &&$<0.01$&& 0.4\er0.2       &0.8\er0.4       \\[-0.5ex]
$f_0(1770)$    &&$<0.01$&& 0.2\er0.2     &0.5\er0.2         \\[-1.2ex]
\tiny both       &&&&\tiny $0.5^{+1.1}_{-0.1}$        & \\ \hline
$f_0(2020)$    &&0.06\er0.03&& 1.7\er0.6       &3.1\er0.8
\\ \hline $f_0(2100)$    &&0.03\er0.02&& 0.6\er0.3       &1.4\er0.5
\\ \hline $f_0(2200)$    &&0.02\er0.02&&  0.7\er0.4       &1.0\er0.6
\\[0.5ex] \hline\hline $\sum f_0(1710)-f_0(2100)$&& 0.10\er0.03&& 2.9\er0.4 &  4.9\er0.8\\[0.5ex] \hline\hline
$f_2(1270)$    &&0.025\er0.010&& 0.10\er0.05        & $<0.05$      \\[-1.2ex]
                     &&\tiny 0.33$^{+0.5}_{-0.6}$&& \tiny0.11\er0.04        &        \\ \hline
$f_2'(1525)$   &&$\sim 0$&& 0.2\er 0.1   & 12 \er2        \\[-2ex]
                     &&&&                  &\tiny 10.7\er2.4          \\[-0.5ex]\hline
$f_2(1640)$    &&$\sim 0$&&  $\sim0$       &  $\sim0$        \\\hline
$f_2(1810)$    &&$\sim 0$&&  $\sim0$         & 0.2\er0.1      \\\hline\hline
$\rho(770)$    && 2.55*  && &\\[-1.2ex]
                    &&\tiny 2.55$^{+0.18}_{-0.16}$ &&&\\[-0.1ex]\hline
$\rho(1450)$  && 0.2\er0.1  && &\\[-1.5ex]
                    &&\tiny 0.29$^{+0.16}_{-0.07}$ &&&\\\hline
$\rho(1700)$  &&  & &&\\[-1.5ex]
                    &&\tiny 0.20\er0.13& &&\\[-0.5ex]
\hline\hline\vspace{1mm}
\end{tabular}}
\end{table}

Before comparing the BR of the $B_{(s)}\to J/\psi f_0$ decays with those of the
$ J/\psi \to \gamma f_0$ decays, we test the consistency of the two datasets and analyzes.
Table~\ref{ratio} compares the $f_0\to K_sK_s$ and $f_0\to\pi^+\pi^-$ BRs when the
scalar mesons are produced in the reactions 
$B_{(s)}\to J/\psi f_0$ or 
reaction $ J/\psi \to \gamma f_0$. The ratios have rather large uncertainties,
but within the uncertainties, the
results are fully compatible.
\begin{table}[pt]
\tbl{\label{ratio}{$K_s K_s/\pi^+\pi^-$ ratio} of scalar mesons produced in
 $B_{(s)}\to J/\psi f_0$ or reaction $J/\psi\to \gamma f_0$.}
{
\centering
\renewcommand{\arraystretch}{1.4}
\begin{tabular}{cccccccc}
\hline\hline
  State     &  $B^0_{s}$ decays & $J/\psi$ decays &\phantom{rrr}&  State  &  $B^0_{s}$ decays & $J/\psi$ decays  \\ \hline
$f_0(980)$  &0.33\er 0.05 &0.61\er0.23    &&
$f_0(1370)$ &0.21\er 0.07 & 0.34 \er0.14   \\ \hline
$f_0(1500)$ &0.31\er0.17  & 0.33 \er0.13  &&
$f_0(1710)$ &2.7\er 1.9   & 3.8\er1.8   \\ \hline 
$f_0(1770)$ &3.3\er2.1    & 2.5\er1.2     &&
$f_0(2020)$ &2.4\er1.0    & 1.3\er 0.7  \\ \hline 
$f_0(2100)$ &3.1\er1.9    & 1.6\er1.2     &&
$f_0(2200)$ &1.9\er1.6    & 1.0\er1.0\\  \hline\hline
\end{tabular}}
\end{table}

Finally, in Table~\ref{glue} we compare the production rates of scalar and tensor
mesons in decays $B_{s}\to J/\psi f_0, f_2$. 
The table presents only the sum of yields in their $\pi\pi$ and $K\bar{K}$ decay modes.
In both reactions, the initial state couples to the singlet component of the scalar meson wave function.
However, the initial $s\bar{s}$ state excites the singlet component only via one-third of the wave function, while
two gluons couple to $u\bar{u}$, $d\bar{d}$, and $s\bar{s}$. We also provide the ratio
$R_{gg/s\bar{s}}$ of gluon-gluon to $s\bar{s}$ induced meson production. 

The decays $B\to J/\psi f_0, f_2$ are highly suppressed, even for resonances with a small glueball component only. 
With the present precision, a comparison of these decays with those of radiative $J/\psi$ decays is not significant.   

The low BR of $B_{s}\to J/\psi f_0(500)$ indicates that the $f_0(500)$ wave function is predominantly composed of a $n\bar{n}$ or $\pi\pi$ molecular component. This observation contrasts with interpretations suggesting a primarily singlet structure for $f_0(500)$ and an octet structure for $f_0(980)$~\cite{Klempt:2021nuf,Oller:2003vf}. The strong yield of $f_0(980)$ in $B$ and $B_{s}$ decays indicates significant $n\bar{n}$ and $s\bar{s}$ components in its wave function. The pronounced yield of $f_0(500)$ and the lesser yield of $f_0(980)$ in radiative $J/\psi$ decays support the hypothesis of mainly singlet (octet) assignments for $f_0(500)$ and $f_0(980)$, respectively. The combined production rates of these two lowest-mass scalar isoscalar mesons are higher in radiative $J/\psi$ decays by a factor of $3.6\pm0.8$. By summing the contributions of $f_0(500)$ and $f_0(980)$, we mitigate uncertainties in their SU(3) structure: in the sum over both isosinglet mesons, there is (in probability) one-half $u\bar{u}$ and one-half $d\bar{d}$ pairs, along with one $s\bar{s}$ pair in any multiplet. For $f_0(1370)$ and $f_0(1500)$, the total yield is $(15\pm3)\cdot 10^{-5}$ in $B_{s}\to J/\psi f_0$ and $(63\pm11)\cdot 10^{-5}$ in radiative $J/\psi$, with a ratio of $4.2\pm1.1$. For the highest-mass scalar meson, $f_0(2200)$, the ratio is $3.3\pm2.4$.

In the intermediate range, which spans from $f_0(1710)$ to $f_0(2100)$, the total yield in radiative $ J/\psi $ decays amounts to $(262 \pm 24) \cdot 10^{-5}$, compared to $(16 \pm 3) \cdot 10^{-5}$ for $B_s\to J/\psi f_0$. This mass range is particularly important as it is where the scalar
glueball is expected. Here, the production of scalar mesons is significantly higher in radiative $ J/\psi $ decays compared to $B_{s}\to J/\psi f_0$ decays. 

\begin{table}[pt]
\tbl{\label{glue}$B_s$ and $ J/\psi $ decay rates into scalar mesons. The sum of $f_0$
decays into $\pi\pi$ and $K\bar K$ is given (in units of $10^{-5}$)
and the ratio R$_{gg/s\bar s}$.}
{
\renewcommand{\arraystretch}{1.4}
\begin{tabular}{ccccccc}
\hline\hline
&   $BR_{ J/\psi \to\gamma f_0}$  &  $BR_{B_s\to J/\psi   f_0}$ &   R$_{gg/s\bar s}$   \\ \hline
\hline $ f_0(500)$        &  110\er 22  &$<0.4$    &  large\\
\hline $f_0(980)$         &2.1\er 0.4   &30\er3  &0.07\er 0.02 \\
\hline $f_0(1370)$        &51\er 11     &9.1\er 1.9 &5.6\er1.2\\
\hline $f_0(1500)$        &  12\er2     &5.9\er2.0 &2.0\er0.8 \\
\hline $f_0(1710)$        &29\er8       &2.2\er0.9   &13$^{+6}_{-5}$ \\
\hline $f_0(1770)$        & 84\er 22    &1.3\er0.5  &65$^{+25}_{-17}$\\
\hline $f_0(2020)$        & 97\er 27    &8.8\er2.0  &11\er3\\
\hline $f_0(2100)$        & 52\er 22    &3.7\er1.2 &14$^{+5}_{-4}$\\
\hline $f_0(2200)$        & 10\er 5     &3.0\er1.6    &3.3$^{+5.7}_{-1.6}$\\[0.5ex] \hline\hline 
$\sum f_0(1710)-f_0(2100)$& 262\er 32   &16\er 3  &  16.4\er 3.7\\[0.5ex] 
\hline\hline
$f_2'(1525)$              &  61\er 6    &24\er4  \\
\hline $f_2(1640)$        &  $\sim 0$   &$\sim 0$  \\
\hline $f_2(1810)$        & $\sim 0$    &0.4\er0.2  \\
\hline $f_2(2210)$        &  35\er 10   & $\sim 0$ \\\hline\hline 
$\sum f_0(1810)-f_0(2210)$& 35\er 10    & 0.4\er 0.2 & large\\[0.5ex] 
\hline\hline
\end{tabular}}
\end{table}

\section{Interpretation}
We now speculate on why the branching ratio for $ J/\psi \to\gamma f_0$ decays is much larger in a region where scalar mesons were shown to have a significant glueball fraction in their wave function~\cite{Sarantsev:2021ein,Klempt:2021wpg}. In decays of $B_s\to J/\psi f_0$, a pair of $s\bar s$ is created  in a volume given by the size of the $J/\psi$. We assume that this $s\bar s$ pair has 
little overlap with the wave function of a highly excited state. Therefore, the production of high-mass mesons can be expected to be small.

In radiative $J/\psi$ decays, gluons are also produced in a small volume. The overlap with the $q\bar q$ part of the wave function is also expected to be small. However, the glueball part of the wave function does not necessarily have a complex structure; its density is could be concentrated at the origin of the wave function. Thus, we hypothesize that mesons containing a glueball fraction can be strongly produced in radiative $J/\psi$ decays. 

\begin{figure}[pt]
\begin{center}
\vspace{-1mm}
\includegraphics[width=0.7\textwidth]{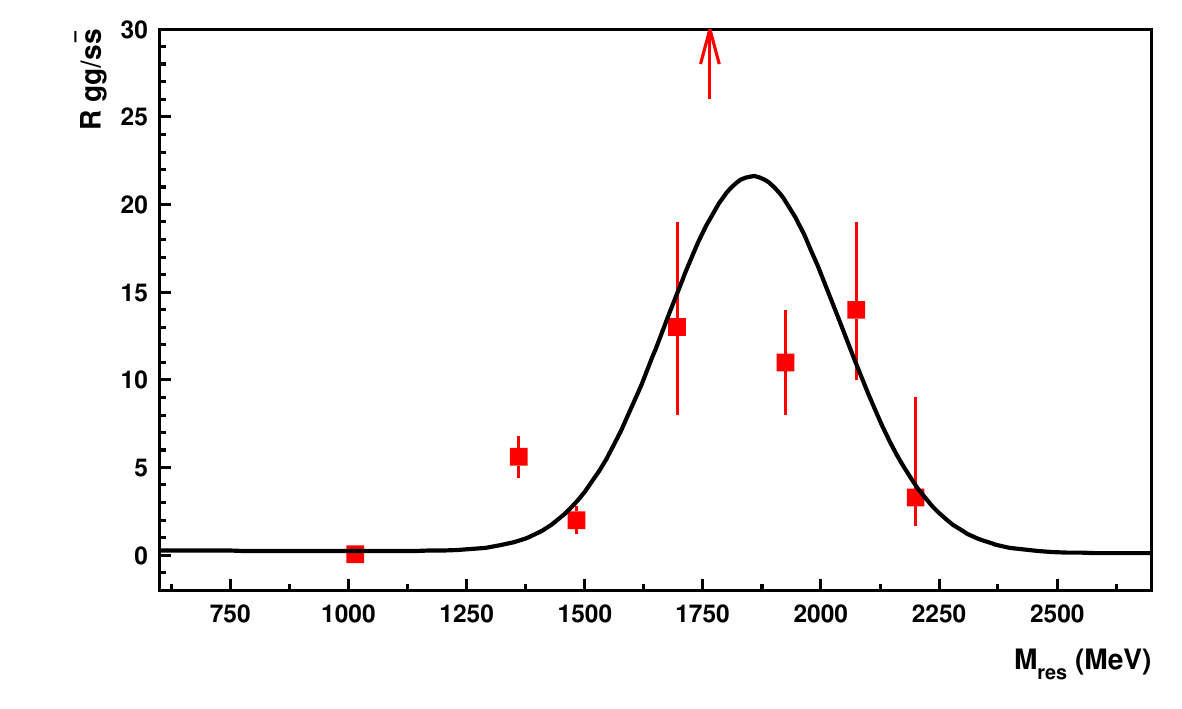}
\vspace{-3mm}
\end{center}
\caption{\label{content}The ratio  R$_{gg/s\bar s}$ of scalar meson production in radiative
$ J/\psi $ decays and in  $B_{(s)}\to J/\psi f_0$ compared to the yield of scalar
mainly-singlet and mainly octet mesons. }
\end{figure}

We denote R${gg/s\bar s}$ as the ratio of the frequency of scalar meson production in $ J/\psi $ decays compared to $B{(s)}\to J/\psi f_0$. This ratio is illustrated in Fig.~\ref{content}. In both cases, the decay of the scalar meson into $\pi\pi$ and $K\bar K$ is considered. We observe that this ratio is enhanced in the region where the scalar glueball is expected. R$_{gg/s\bar s}$ is compared to a Breit-Wigner function with mass and width determined in Ref.\cite{Sarantsev:2021ein}, adjusted for suitable amplitude.

Minor contributions from $B_{(s)}$ decays due to reaction in Fig.~1d could contribute to R$_{gg/s\bar s}$.
Such contributions would make the true ratio R$_{gg/s\bar s}$ even larger.

The ratios R${gg/s\bar s}$, albeit with significant uncertainties, exhibit a similarity to the expected shape of the scalar glueball. However, the ratio for $f_0(1370)$ is notably large; as an SU(3) mainly-singlet and ground state meson, its radiative yield is naturally high. The ratios for $f_0(1710)$ and $f_0(1770)$ deviate from the curve, suggesting that our fit may assign too little intensity to $f_0(1710)$ and too much to $f_0(1770)$ production in the LHCb data. The mean value for both yields $31^{+19}{-14}$, which is consistent with the expected curve. It is important to note that with low statistics, accurately attributing fractional contributions to overlapping resonances remains challenging.

In $B_s\to J/\psi f_2$ decays, we observe practically no contributions above $f_2'(1525)$. In contrast, in radiative $ J/\psi$ decays into $\pi\pi$, a significant intensity is found at 2210,MeV~\cite{Klempt:2022qjf}. The low-mass tensor mesons are nearly ideally mixed, suggesting that a high-mass $n\bar n$ tensor meson may also be ideally mixed. Consequently, it would be produced only via the $d\bar d$ initial state and not via $s\bar s$.

In $B\to J/\psi +$ meson decays, the phase space ends at 2183,MeV. Therefore, the 2210,MeV peak position cannot be directly observed, but 
there is no indication of a resonance onset at the end of the phase space. Above $f_2(1270)$, we observe practically no tensor contribution.
Similarly, in radiative $J/\psi$ decays, the intensity above $f_2(1270)$ or $f_2'(1525)$ is also rather low. 
However, the intensity starts to increase only above 1900,MeV and peaks at 2210,MeV.

We interpret the enhancements of intensity in radiative $ J/\psi $ decays at 1865,MeV in the scalar wave and at 2210,MeV in the tensor wave as effects of gluon-gluon interactions. The wave function of highly excited scalar states has little overlap with the initial $s\bar s$ state, resulting in a small production rate. This is particularly evident in the case of tensor-meson production.

In radiative $ J/\psi $ decays, the two initial-state gluons can form a glueball through mixing with scalar or tensor mesons. Unlike direct production from $s\bar s$, this mixing is not suppressed by wave function effects. Therefore, the production rate includes a component that is proportional to the glueball content. The high production yields of the scalar and tensor mesons compared to those in $B_{(s)}\to J/\psi +$ meson decays support
the assumption that the scalar and the tensor glueball play a significant role in radiative $J/\psi$ decays.
Note that these conclusions remain unchanged when the data on radiative $J/\psi$ decays are fitted with a smaller number
of resonances as in~\cite{Rodas:2021tyb}. Relevant is the total intensity in the mass range from 1600 to 2200 and not how
the intensity is shared between individual resonances.

\section{\label{sum}
Summary}
We have fitted the LHCb data on $B^0\to J/\psi  \pi\pi$, $B_{s} ^0\to J/\psi  \pi^+\pi^-$, and
$B_{s} ^0\to J/\psi  K^+ K^-$
and compared the branching ratios for the production of scalar and tensor mesons
with those observed in radiative $J/\psi$ decays. Above 1500\,MeV, only little intensity is observed in
$B_{(s)}^0\to J/\psi $ $f_0,f_2$ decays, while radiative $ J/\psi $ decays show significant intensities
at higher mass, peaking at 1865\,MeV in the scalar and at 2210\,MeV in the tensor wave.
We argue that the little intensity in $B_{(s)}\to J/\psi  f_0,f_2$ decays is due to the small
overlap of the wave function of highly excited scalar or tensor mesons with the initial
$d\bar d$ or $s\bar s$ pair while in radiative decay, the initial two gluons form a glueball
that mixes with the wave function of excited states, and the production of mesons
with masses close to a glueball mass is enhanced. The LHCb data thus support the glueball
interpretation of $G_0(1865)$ and $G_2(2210)$.

\section*{Acknowledgement}
Funded by the Deutsche Forschungsgemeinschaft (DFG, German Research
Foundation) – Project-ID 196253076 – TRR 110T.

\end{document}